\documentclass[english]{article}
\usepackage{times}
\usepackage[T1]{fontenc}
\usepackage{geometry}
\usepackage{amsmath}
\usepackage{amssymb}

\makeatletter

\providecommand{\LyX}{L\kern-.1667em\lower.25em\hbox{Y}\kern-.125emX\@}

\newcommand{\sd}{\, / \! \! \! \! \nabla}

\usepackage{babel}
\makeatother
\begin{document}

\title{Clifford Geometrodynamics}

\author{A. Garrett Lisi%
\footnote{Gar@Lisi.com%
}}

\maketitle
\begin{abstract}
Classical anti-commuting spinor fields and their dynamics are derived
from the geometry of the Clifford bundle over spacetime via the BRST
formulation. In conjunction with Kaluza-Klein theory, this results
in a geometric description of all the fields and dynamics of the standard
model coupled to gravity and provides the starting point for a new
approach to quantum gravity.
\end{abstract}

\section{Introduction}

In most approaches to the classical and quantum dynamics of spinor
fields the fields and their properties are postulated ad-hoc, without
any geometric motivation as to why they should exist, but only the
rationale that they are necessary to represent fermions. The desired
mathematical structures, such as complex valued matrix columns and
the necessity that the field components anti-commute, are put in by
hand along with the equations of motion, without any geometric justification.
It is the purpose of this paper to propose a new geometric foundation
and justification for the existence and dynamics of spinor fields.
In so doing, all the fields and dynamics of the standard model may
be derived from pure geometry.

The key to this construction is to begin with the Clifford algebra
bundle (or the associated matrix bundle) as the fundamental geometric
framework. This fibre bundle has a connection and curvature, and a
frame provides a bundle map to the cotangent bundle and gives the
metric on the pseudo-Riemannian base manifold. The dynamics of the
Clifford connection and frame is given by extremizing the total scalar
Clifford bundle curvature, the gravitational action \cite{peldan}.
The curvature is invariant under adjoint automorphisms of the Clifford
bundle, and this gauge symmetry, when properly restricted using the
BRST formulation \cite{holten}, results in the appearance and familiar
dynamics of a pair of anti-commuting, Clifford valued spinor fields.
The result is the coupled system of gravitational and spinor fields.
A Kaluza-Klein decomposition of the vielbein then provides the vector
gauge fields, completing the picture.

\section{Geometric framework}

To describe the geometry of the physical universe, an $n$ dimensional,
pseudo-Riemannian differential manifold is used as the base space
for the Clifford algebra fibre bundle. The $n$ basis vector elements,
$\left\{ \gamma ^{\alpha }\right\} $, for the Clifford bundle provide
a local trivialization, $\partial _{i}\gamma ^{\alpha }=\frac{\partial }{\partial x^{i}}\gamma ^{\alpha }=0$,
and satisfy, under the symmetric product,\begin{equation}
\gamma ^{\alpha }\bullet \gamma ^{\beta }=\frac{1}{2}\left(\gamma ^{\alpha }\gamma ^{\beta }+\gamma ^{\beta }\gamma ^{\alpha }\right)=\eta ^{\alpha \beta }\label{eq:funca}\end{equation}
The Clifford algebra, $Cl$, has a faithful representation in the
complex matrices, $GL(2^{[n/2]},C)$, with the Clifford product isomorphic
to matrix multiplication, and it is possible and sometimes helpful,
although not necessary, to write and manipulate Clifford elements
as matrices, using (Dirac) matrices for the basis.

The fundamental Clifford algebra equation (\ref{eq:funca}) is invariant
under the Clifford group, $Cl^{\star }=\left\{ S\in Cl\mid \exists S^{-1}\right\} $,
adjoint automorphism (similarity transformation), $\gamma ^{\alpha }\mapsto \gamma '{}^{\alpha }=S\gamma ^{\alpha }S^{-1}$,
which may well change the grade of the Clifford basis elements. Through
this automorphism the $Cl^{\star }$ elements serve as the transition
functions for the Clifford bundle. For infinitesimal transformations,
$S\simeq 1+\frac{1}{2}C$, $C\in Cl$, this automorphism is\begin{equation}
\gamma '{}^{\alpha }=S\gamma ^{\alpha }S^{-1}\simeq \gamma ^{\alpha }+C\times \gamma ^{\alpha }\label{eq:adj}\end{equation}
with the anti-symmetric (cross) product defined as $A\times B=\left[A,B\right]=\frac{1}{2}\left(AB-BA\right)$.

In general, the Clifford product may be written as $AB=A\bullet B+A\times B$
and, since $\gamma ^{\alpha }\gamma ^{\beta }=\eta ^{\alpha \beta }+\gamma ^{\alpha }\times \gamma ^{\beta }$,
any Clifford element may be written out with real, anti-symmetric
coefficients times anti-symmetric collections of the basis elements,\[
C=C_{s}+C_{\alpha }\gamma ^{\alpha }+C_{\alpha \beta }\left[\gamma ^{\alpha },\gamma ^{\beta }\right]+C_{\alpha \beta \delta }\left[\gamma ^{\alpha },\gamma ^{\beta },\gamma ^{\delta }\right]+...+C_{p}\gamma \]
where the anti-symmetric bracket operator is, for example,\[
\begin{array}{rcl}
 \left[A,B,C\right] & = & \frac{1}{3!}\left(ABC+BCA+CAB-ACB-CBA-BAC\right)\\
 C_{\alpha \beta \delta } & = & C_{\left[\alpha \beta \delta \right]}=\frac{1}{3!}\left(C_{\alpha \beta \delta }+C_{\beta \delta \alpha }+C_{\delta \alpha \beta }-C_{\alpha \delta \beta }-C_{\delta \beta \alpha }-C_{\beta \alpha \delta }\right)\end{array}
\]
Note that, since the basis elements can change grade under the adjoint
automorphism, the grade of a Clifford element, except for the invariant
scalar part (grade $0$ part, trace), is only a meaningful concept
with respect to the local specification of the basis elements. The
$\cdot $ (dot) and $\wedge $ (wedge) products familiar to disciples
of Clifford algebra have grade dependent definitions and are therefore
not used here, though they are equivalent to $\bullet $ and $\times $
for vector elements and useful within the context of a local vector
basis.

The connection for the $Cl$ bundle takes values in the Lie algebra
of the transitions and is thus a Clifford element acting via the cross
product, allowing the covariant derivative of the basis elements to
be written as\[
\nabla _{i}\gamma ^{\alpha }=\Omega _{i}\times \gamma ^{\alpha }\]
giving, for any Clifford element, $\nabla _{i}C=\partial _{i}C+\Omega _{i}\times C$.
Under an adjoint automorphism (\ref{eq:adj}) this connection transforms
as\begin{equation}
\Omega _{i}\mapsto \Omega '_{i}=2S\partial _{i}S^{-1}+S\Omega _{i}S^{-1}=2S\sd _{i}S^{-1}\simeq \Omega _{i}-\nabla _{i}C\label{eq:contran}\end{equation}
in which the Dirac derivative operator is introduced as $\sd _{i}=\partial _{i}+\frac{1}{2}\Omega _{i}$.

The Clifford basis is related to the metric via the fundamental frame
(frame, soldering form, bundle map),\begin{equation}
\begin{array}{rcl}
 \hat{e} & = & \gamma ^{\alpha }\left(e_{\alpha }\right)^{i}\vec{\partial }_{i}=\gamma ^{\alpha }\hat{e}_{\alpha }=\gamma ^{i}\vec{\partial }_{i}\\
 \underline{e} & = & \underrightarrow{dx^{i}}\left(e^{-1}{}_{i}\right)^{\alpha }\gamma _{\alpha }\end{array}
\label{eq:frame}\end{equation}
which gives the metric through the relation for the orthonormal basis
(vielbein, tetrad, ONB), $g_{ij}=\left(e^{-1}{}_{i}\right)^{\alpha }\eta _{\alpha \beta }\left(e^{-1}{}_{j}\right)^{\beta }$.

\section{Dynamics}

The curvature of the Clifford bundle, giving the parallel transport
of Clifford elements around infinitesimal loops, is the $Cl$ valued
$2$-form,\begin{equation}
\begin{array}{rcl}
 \underrightarrow{\underrightarrow{R}} & = & \underrightarrow{dx^{i}}\, \underrightarrow{dx^{j}}\frac{1}{2}R_{ij}=\underrightarrow{d}\, \underrightarrow{\Omega }+\frac{1}{2}\underrightarrow{\Omega }\times \underrightarrow{\Omega }=\underrightarrow{d}\, \underrightarrow{\Omega }+\frac{1}{2}\underrightarrow{\Omega }\, \underrightarrow{\Omega }=\underrightarrow{\sd }\, \underrightarrow{\Omega }\\
 R_{ij} & = & \partial _{i}\Omega _{j}-\partial _{j}\Omega _{i}+\Omega _{i}\times \Omega _{j}\end{array}
\label{eq:r2}\end{equation}
The scalar curvature of the Clifford bundle may be defined, using
the frame (\ref{eq:frame}), as\begin{equation}
R=\left\langle \underrightarrow{\underrightarrow{R}}\hat{e}\hat{e}\right\rangle =\left\langle R_{ij}\left(e_{\alpha }\right)^{i}\left(e_{\beta }\right)^{j}\gamma ^{\alpha }\gamma ^{\beta }\right\rangle \label{eq:r}\end{equation}
with $\left\langle A\right\rangle $ giving the scalar part (trace)
of a Clifford element.

The action for the Clifford bundle is\begin{equation}
S=\int \underrightarrow{d^{n}x}\left|e\right|R=\int \underrightarrow{d^{n}x}\left|e\right|\left\langle 2\left(\partial _{i}\Omega _{j}+\frac{1}{2}\Omega _{i}\Omega _{j}\right)\left(e_{\alpha }\right)^{i}\left(e_{\beta }\right)^{j}\left(\gamma ^{\alpha }\times \gamma ^{\beta }\right)\right\rangle \label{eq:action}\end{equation}
with volume scale $\left|e\right|=\det \left(e^{-1}{}_{i}\right)^{\alpha }$.
Varying the vielbein and requiring the variation of this action to
vanish gives Einstein's equation,\[
0=G_{i}{}^{\alpha }=\left\langle \left(\partial _{\left[i\right.}\Omega _{\left.j\right]}+\frac{1}{2}\Omega _{\left[i\right.}\Omega _{\left.j\right]}\right)\left(e_{\beta }\right)^{j}\left(\gamma ^{\alpha }\times \gamma ^{\beta }\right)\right\rangle +\frac{1}{2}\left(e^{-1}{}_{i}\right)^{\alpha }R\]
and varying the Clifford connection gives a relationship with derivatives
of the vielbein\[
\left[\partial _{i}\left|e\right|\left(e_{\alpha }\right)^{i}\left(e_{\beta }\right)^{j}\right]\left(\gamma ^{\alpha }\times \gamma ^{\beta }\right)=\left|e\right|\left(e_{\alpha }\right)^{i}\left(e_{\beta }\right)^{j}\left(\gamma ^{\alpha }\times \gamma ^{\beta }\right)\times \Omega _{i}\]
which holds if and only if the Clifford connection is equal to the
torsionless spin connection bivector,\[
\Omega _{i}=\Omega _{i\alpha \beta }\left[\gamma ^{\alpha },\gamma ^{\beta }\right]\; \; ,\; \; \Omega _{i\alpha \beta }=\omega _{i\alpha \beta }\]
satisfying Cartan's structure equation, $\underrightarrow{d}\, \underline{e}{}^{\alpha }=\underrightarrow{\omega }{}_{\beta }{}^{\alpha }\underline{e}{}^{\beta }$,
equivalent to $\partial _{\left[i\right.}\left(e^{-1}{}_{\left.j\right]}\right)^{\alpha }=\omega _{\left[i\right.\beta }{}^{\alpha }\left(e^{-1}{}_{\left.j\right]}\right)^{\beta }$.

\section{BRST}

The curvature scalar, and thus the action (\ref{eq:action}), is invariant
under adjoint automorphisms of the frame,\[
\begin{array}{rcl}
 \hat{e} & \mapsto  & \hat{e'}=S\hat{e}S^{-1}\simeq \hat{e}+C\times \hat{e}\\
 \underrightarrow{\Omega } & \mapsto  & \underrightarrow{\Omega '}=2S\underrightarrow{\sd }S^{-1}\simeq \underrightarrow{\Omega }-\underrightarrow{\nabla }C\end{array}
\Rightarrow R\mapsto R'=R\]
Via the BRST formulation, new {}``gauge ghost'' fields, with real,
anti-commuting coefficients (Grassmann number $1$, satisfying $ab=-ba$
and $\left(ab\right)^{*}=-b^{*}a^{*}=-ba$), are introduced to properly
restrict and account for this symmetry. The new variables are the
anti-commuting $Cl$ element fields, $\left\{ C,B\right\} $, which
have real coefficients with Grassmann number $1$, and another ghost
field, $A$, with Grassmann number $0$. The infinitesimal BRST transformation
corresponding to the Clifford group adjoint operation is\[
\begin{array}{rcl}
 \delta _{\Lambda }\hat{e} & = & C\times \hat{e}=\frac{1}{2}C\hat{e}-\frac{1}{2}\hat{e}C\\
 \delta _{\Lambda }\underrightarrow{\Omega } & = & -\underrightarrow{\nabla }C=-\underrightarrow{dx^{i}}\left(\partial _{i}C+\frac{1}{2}\Omega _{i}C-\frac{1}{2}C\Omega _{i}\right)\\
 \delta _{\Lambda }C & = & \frac{1}{2}C\times C=\frac{1}{2}CC\\
 \delta _{\Lambda }B & = & iA\\
 \delta _{\Lambda }A & = & 0\end{array}
\]
The nilpotence of the BRST operator, $\delta _{\Lambda }\delta _{\Lambda }=0$,
which has Grassmann number $1$, is confirmed by calculation,\[
\begin{array}{rcl}
 \delta _{\Lambda }\delta _{\Lambda }\hat{e} & = & \frac{1}{2}\left[\frac{1}{2}CC\right]\hat{e}-\frac{1}{2}C\left[\frac{1}{2}C\hat{e}-\frac{1}{2}\hat{e}C\right]-\frac{1}{2}\left[\frac{1}{2}C\hat{e}-\frac{1}{2}\hat{e}C\right]C-\frac{1}{2}\hat{e}\left[\frac{1}{2}CC\right]=0\\
 \delta _{\Lambda }\delta _{\Lambda }\underrightarrow{\Omega } & = & -\underrightarrow{dx^{i}}\left(\partial _{i}\left[\frac{1}{2}CC\right]-\frac{1}{2}\left[\partial _{i}C+\Omega _{i}\times C\right]C+\frac{1}{2}\Omega _{i}\left[\frac{1}{2}CC\right]-\frac{1}{2}\left[\frac{1}{2}CC\right]\Omega _{i}-\frac{1}{2}C\left[\partial _{i}C+\Omega _{i}\times C\right]\right)=0\\
 \delta _{\Lambda }\delta _{\Lambda }C & = & \frac{1}{2}\left[\frac{1}{2}CC\right]C-\frac{1}{2}C\left[\frac{1}{2}CC\right]=0\end{array}
\]
The dynamics of the gauge and ghost degrees of freedom are determined
by the choice of a BRST potential; a good choice is\[
\Psi =\int \underrightarrow{d^{n}x}\left|e\right|\left\langle B\hat{e}\underrightarrow{\Omega }\right\rangle \]
which gives the new action,\[
S{}'=S-i\delta _{\Lambda }\Psi =\int \underrightarrow{d^{n}x}\left|e\right|\left\{ R\left[\hat{e},\, \underrightarrow{\Omega }\right]+\left\langle A\hat{e}\underrightarrow{\Omega }\right\rangle -i\left\langle B\left[\gamma ^{\alpha }\left(e_{\alpha }\right)^{i}\partial _{i}C+\left(\hat{e}\underrightarrow{\Omega }\right)\times C\right]\right\rangle \right\} \]
The ghost field $A$ appears in this action as a Lagrange multiplier,
constraining the connection to satisfy $\hat{e}\underrightarrow{\Omega '}=0$.
With this restricted connection the effective action for the remaining
fields, $\left\{ \hat{e},\underrightarrow{\Omega '},C,B\right\} $,
is\begin{equation}
S_{eff}=\int \underrightarrow{d^{n}x}\left|e\right|\left\{ R\left[\hat{e},\underrightarrow{\Omega '}\right]-i\left\langle B\gamma ^{\alpha }\left(e_{\alpha }\right)^{i}\partial _{i}C\right\rangle \right\} \label{eq:effs}\end{equation}
an Einstein-Weyl-like action for anti-commuting spinor field, anti-field,
vielbein, and restricted connection. Note that the constraint on the
connection, a result of the choice of BRST potential, insures that
the connection vanishes from the Dirac operator. 

The equations of motion from the new action, $S{}'$, or, after removing
$A$ and restricting to $\Omega '_{i}$, from $S_{eff}$, are\[
\begin{array}{rcl}
 G'_{i}{}^{\alpha } & = & \frac{i}{2}\left\langle B\gamma ^{\alpha }\partial _{i}C\right\rangle -\frac{i}{2}\left(e^{-1}{}_{i}\right)^{\alpha }\left\langle B\gamma ^{\beta }\left(e_{\beta }\right)^{j}\partial _{j}C\right\rangle =T_{i}{}^{\alpha }\\
 A & = & -2\frac{\left(n-2\right)}{\left(n-1\right)}\left|e\right|\omega ^{\beta }{}_{\alpha \beta }\gamma ^{\alpha }-4\left|e\right|\omega _{\left[\alpha \beta \delta \right]}\left[\gamma ^{\alpha },\, \gamma ^{\beta },\, \gamma ^{\delta }\right]\\
 \Omega '_{\delta } & = & \left(e_{\delta }\right)^{i}\Omega '_{i}=\Omega '_{\delta \alpha \beta }\left[\gamma ^{\alpha },\gamma ^{\beta }\right]\\
 \Omega '_{\delta \alpha \beta } & = & \omega _{\delta \alpha \beta }-\omega _{\left[\delta \alpha \beta \right]}+\frac{2}{\left(n-1\right)}\eta _{\delta \left[\alpha \right.}\omega ^{\gamma }{}_{\left.\beta \right]\gamma }\\
 0 & = & \gamma ^{\alpha }\partial _{i}\left|e\right|B\left(e_{\alpha }\right)^{i}\\
 0 & = & \gamma ^{\alpha }\left(e_{\alpha }\right)^{i}\partial _{i}C\end{array}
\]
The restricted connection, satisfying $\gamma ^{\alpha }\Omega '_{\alpha }=0$,
is hence a Clifford bivector with gauge degrees of freedom removed--the
coefficients constrained to have vanishing trace $\Omega '_{\delta \alpha }{}^{\delta }=0$
and vanishing fully anti-symmetric part $\Omega '_{\left[\delta \alpha \beta \right]}=0$.
Note that if one begins with an arbitrary connection, the restricted
connection may be obtained by applying an adjoint transformation (\ref{eq:contran})
with an $S$ such that $0=\hat{e'}\underrightarrow{\Omega '}=2S\gamma ^{\alpha }\left(e_{\alpha }\right)^{i}\sd _{i}S^{-1}$--so
the geometric interpretation of a solution to the curved spacetime
Weyl equation is that it provides a transformation to a restricted
connection. The BRST formulation balances the gauge restriction with
the $C$ and $B$ fields and their (Weyl) equations of motion and
stress energy tensor. By using the equation for the restricted Clifford
connection in terms of the spin connection it is also possible to
write the Clifford scalar curvature (\ref{eq:r}), and hence a new
effective action, purely in terms of derivatives of the vielbein,\[
\begin{array}{rcl}
 R'\left[\hat{e}\right] & = & \left\langle 2\left(\partial _{i}\Omega '_{j}+\frac{1}{2}\Omega '_{i}\Omega '_{j}\right)\left(e_{\alpha }\right)^{i}\left(e_{\beta }\right)^{j}\left(\gamma ^{\alpha }\times \gamma ^{\beta }\right)\right\rangle \\
  & = & \frac{1}{3}\omega _{\beta \alpha \delta }\omega ^{\alpha \delta \beta }-\frac{1}{3}\omega _{\beta \alpha \delta }\omega ^{\beta \alpha \delta }+\frac{1}{\left(n-1\right)}\omega ^{\delta }{}_{\beta \delta }\omega _{\gamma }{}^{\beta \gamma }\\
  & = & -\frac{2}{3}F_{\delta \left(\beta \alpha \right)}F^{\delta \left(\beta \alpha \right)}+\frac{1}{\left(n-1\right)}F^{\delta }{}_{\beta \delta }F_{\gamma }{}^{\beta \gamma }\end{array}
\]
in which the field strength (anholonomy) for the vielbein is defined
as $F_{\beta \gamma }{}^{\alpha }=\left(e_{\beta }\right)^{i}\left(e_{\gamma }\right)^{j}2\partial _{\left[i\right.}\left(e^{-1}{}_{\left.j\right]}\right)^{\alpha }$.

\section{Conclusion}

The existence and dynamics of anti-commuting spinor fields have been
derived and given a firm geometric foundation by starting with the
Clifford algebra fibre bundle and curvature and applying the BRST
formulation to the adjoint automorphism gauge symmetry. The BRST construction
was carried out in the relativistic Lagrangian framework, but may
be carried through to a Hamiltonian formulation as well.

The ultimate goal is to obtain all the fields and dynamics of the
standard model of particle physics from this geometric foundation.
To do this, the methods of Kaluza-Klein theory may be employed by
assuming the dimensions greater than the four of spacetime are wrapped
up in a spatial, highly symmetric, compact manifold. By using a vielbein
of the form\[
\left(e_{\alpha }\right)^{i}=\left[\begin{array}{cc}
 \left(e_{\alpha }^{S}\right)^{i} & A_{\alpha }^{A}\xi _{A}^{i}\\
 0 & \frac{1}{\rho }\left(e_{\alpha }^{K}\right)^{i}\end{array}
\right]\]
in which $\vec{\xi }_{A}$ are Killing vector fields of the compact
space, $K$, one obtains the Yang-Mills action and dynamics for the
gauge fields, $\underrightarrow{A}{}^{A}$. By expanding these gauge
and spinor fields in terms of resonant modes of the compact space
one may get all the fields and dynamics of the standard model along
with gravity. The details of how these higher dimensional, Clifford
element spinor fields can be broken up into the familiar fermion multiplets
is laid out in a beautiful exposition by Trayling \cite{trayling}.
Note that the spinor fields derived here are not originally relegated
to ideals of the algebra, but rather are {}``full'' Clifford element
(matrix) spinors--the ideals (matrix columns) emerge only in the fermion
multiplet decomposition.

It is important to note that not all the gauge symmetries of the system
have yet been addressed. The other two symmetries, transformations
of the frame that leave the action invariant, are diffeomorphisms
(coordinate changes) and local Lorentz rotations of the vielbein.
These symmetries may also be handled via the BRST formulation, resulting
in an effective action for a restricted vielbein. Used in conjunction
with Kaluza-Klein theory, the BRST formulation for diffeomorphisms
parallels and reproduces the BRST formulation for Yang-Mills gauge
theory, with the relevant diffeomorphism, producing the familiar gauge
field transformation, being $x^{i}\mapsto x'{}^{i}=x^{i}+\xi _{A}^{i}\phi ^{A}\left(x\right)$.

The BRST formulation for Yang-Mills theory, and the appearance and
utility of gauge ghosts, is familiar to researchers in quantum field
theory, where the ghosts play a crucial role in facilitating quantization
and renormalization. It is hoped that the spinor ghost fields and
dynamics introduced here (\ref{eq:effs}) may play a similar role
in the quantization of gravity.

\section{Acknowledgments}

This research was made possible by a generous grant from the Hopgood
Foundation. The typesetting was done using \LyX{}%
\footnote{http://www.lyx.org%
}, an excellent open source program. As he has been a bit of a recluse,
the author would also like to acknowledge many long and fruitful conversations
with himself. However, in the interest of the work, the author heartily
welcomes any corrections and comments that may be forthcoming from
those who wish to contribute them.

\end{document}